\documentclass{aa}
\usepackage{txfonts}
\usepackage{natbib}
\usepackage{graphicx}
\bibpunct{(}{)}{;}{a}{}{,} % to follow the A&A style
\usepackage{longtable}
\usepackage{xspace}
\usepackage{longtable}
\usepackage{epsfig}
\usepackage{verbatim}

\newcommand\micron{\mbox{$\mu$m}\xspace}
\newcommand\ecsaa{erg~cm$^{-2}$~s$^{-1}$~\AA$^{-1}$~arcsec$^{-2}$\xspace}
\newcommand\msun{$M_{\sun}$\xspace}

\begin{document}

\title{Observation of light echoes around very young stars}

\author{Ortiz, J.L. \inst{1}
\and Sugerman, B.E.K. \inst{2} \and de la Cueva, I. \inst{3} \and
Santos-Sanz, P.\inst{1,4} \and Duffard, R. \inst{1} \and Gil-Hutton, R.
\inst{5} \and Melita, M. \inst{6} \and Morales, N. \inst{1}}

% \offprints{Duffard, R. \email{duffard@iaa.es}}

\institute{Instituto de Astrof\'{\i}sica  de Andaluc\'{\i}a - CSIC, Apt
3004, 18080  Granada,  Spain \and Department of Physics and Astronomy,
Goucher College, 1021 Dulaney Valley Rd., Baltimore, Maryland 21204, USA.
\and Astroimagen. C. Abad y Lasierra, 58 Bis - 62, 07800 Ibiza, Islas
Baleares, Spain. \and Observatoire de Paris, LESIA-UMR CNRS 8109, 5 place
Jules Janssen F-92195 Meudon cedex, France. \and Complejo Astron\'omico El
Leoncito (CASLEO-CONICET) and San Juan National University, Avda. de
Espa\~{n}a 1512 sur, J5402DSP, San Juan, Argentina. \and Instituto de
Astronom\'{\i}a y F\'{\i}sica del Espacio, Universidad de Buenos Aires,
CONICET, CC 67-Suc 28, C1428ZAA, Ciudad Aut\'{o}noma de Buenos Aires,
Argentina.}

\titlerunning{Light Echoes from S CrA and R CrA}

\abstract{} {The goal of the paper is to present new results on light echoes
from young stellar objects.} {Broad band CCD images were obtained over three
months at one-to-two week intervals for the field of NGC 6726, using the
large field-of-view remotely-operated telescope on top of Cerro Burek.}{We
detected scattered light echoes around two young, low-amplitude, irregular
variable stars. Observations revealed not just one, but multiple light
echoes from brightness pulses of the T Tauri star S CrA and the Herbig Ae/Be
star R CrA. Analysis of S CrA's recurring echoes suggests that the star is
located $138 \pm 16$ pc from Earth, making these the closest echoes ever
detected. The environment that scatters the stellar light from S CrA is
compatible with an incomplete dust shell or an inclined torus some 10,000 AU
in radius and containing $\sim$ $2 \times 10^{-3}$ $M_{\sun}$ of dust. The
cause of such concentration at $\sim$ 10,000AU from the star is unknown. It
could be the remnant of the envelope from which the star formed,  but the
distance of the cloud is remarkably similar to the nominal distance of the
Oort cloud to the Sun, leading us to also speculate that the dust (or ice)
seen around S CrA might have the same origin as the Solar System Oort
cloud.}{}
 \maketitle
%
%________________________________________________________________

\section{Introduction}

Light echoes can be produced when the light pulse from an astronomical
source is scattered by dust. If the direction of the scattering is toward
the Earth and the pulse is intense enough, an echo can be observed from
Earth \citep{Couderc1939}, provided that the number density of dust
particles is high enough. All those conditions are not usually met. In fact,
light echoes are impressive and very rare phenomena that have been witnessed
in only a handful of events \citep[][and references
therein]{Couderc1939,Suntzeff1988,Bode1985,Rest2005,Krause2005,Bond2003,Suger2005,Quinn2006,Suger2003,Crotts2008,Liu2003,Rest2008a,Rest2008b,Cappe2001,Sparks1999,Sparks2008,Bond2009}.
Such events are mostly explosive or nearly explosive (supernovae, novae,
novae-like, a Cepheid variable, and a planetary nebula) from stellar objects
in their latest stages of evolution.

Since scattered-light echoes reveal the 3-dimensional structure and chemical
composition of dusty circumstellar and interstellar environments around
variable sources, they should serve as powerful probes of early stages of
star and planet formation, when the young stellar sources exhibit irregular,
low-amplitude variability. Thus, finding and analyzing echoes from young
stellar sources would be a powerful tool to study stellar and planetary
formation. Although theoretical predictions exist for the detectability of
echoes around young variable stars \citep[e.g.][]{Suger2003,Gaidos1994},
systematic searches have not been carried out, perhaps because the low
detection rate from intentional searches around supernovae \citep{Boffi1999}
has suggested that fainter echoes are especially difficult to observe.

However, with current and suitable technology, faint echoes from young stars
can be observed. Here we present the detection of multiple light echoes from
two young variable stars, the Herbig Ae/Be star R CrA and T Tauri star S
CrA. It must be mentioned that for R CrA, short-term variability in the
brightness of its nebulosity (NGC 6729) has been reported before (e.g.
Hubble 1921, Graham and Phillips 1989), and also similar changes have been
reported in other variable nebulae like NGC 2261
\citep[e.g.][]{Hubble1921,Lightfoot1989} and in NGC 1555. The question of
light variability in nebulae was a key issue at the end of the 19th century,
and there are reports going back to 1862 \citep{Struve1862} about the
variability of NGC 1555. A somewhat newer account of light variability in
the complex surroundings of T Tauri is given by \cite{Barnard1895}.

After a careful reading of Hubble reports (1921) on several variable
nebulae, a phenomenon of the expansion of a bright envelope around R CrA (of
just a few arcsecs in a few days), which is called in the report ``...a wave
of illumination moving..." qualifies as a true light echo, although it was
not realized as such at the time. On the other hand, the derivation of the
distance to the star made by Hubble was based on incorrect analysis, like
the analyses of the Nova Per (1901) light echo prior to Couderc's (1939)
correct interpretation of the light echo phenomenon. Maybe those are the
reasons the R CrA inner core expansion phenomenon has never been recognized
in the light echo literature so is not listed in any light echo compilation.
Regardless of when the first detection of a light echo around a young star
took place, it must be pointed out that the R CrA light echoes presented
here are much farther away from the star than ``the wave of illumination"
reported in Hubble (1921) and farther away than the variability areas
reported in previous works \citep[e.g.][]{Hubble1921,Belli1980,Graham1987}.
Concerning the T Tauri S CrA, no variability in its nebula has ever been
reported and the finding of its light echo is reported here for the first
time. In this paper we focus on a preliminary analysis of the echoes
surrounding  S CrA.

\section{Observations and Reductions}

As part of a southern-hemisphere sky survey carried out at the Complejo
Astron\'omico El Leoncito (CASLEO) in Argentina, a field containing the
reflection nebula NGC 6726 was imaged twice, with roughly one month between
observations. The observations were made by means of our large field-of-view
0.45m f/2.8 remotely-operated telescope on top of Cerro Burek, through a
broad (390 to 700 nm) filter. The detector was a 4008 $\times$ 2672 pixel
camera based on the Kodak KAF11000M CCD chip. The pixel scale of the setup
was 1.47 arcsec/pixel. In the course of standard reduction and analysis, the
very young stars S CrA and R CrA experienced significant brightness changes.
Closer inspection of the data clearly revealed material that appeared to be
expanding away from the two stars at superluminal speeds for all plausible
values of the distances to the stars.  The appearance of such superluminal
motion is well understood as the scattering of a light pulse off of
stationary dust near the line of sight, rather than physical expansion of
material ejected by the star \citep{Couderc1939}.  Light echoes not only
illuminate otherwise hidden dust, but also directly reveal that material's
3-dimensional structure.  Since mapping the circumstellar environments of
such young stars has the potential to reveal critical details of their
formation processes, we re-imaged the field of NGC 6726  over the next three
months  at one-to-two week intervals, as listed in Table 1.  Each image was
reduced in the standard manner (e.g.\ bias subtraction, flat fielding), then
geometrically registered to a common orientation using field stars as
described in \citet{Suger2005}. RMS residuals of geometric registration are
less than 0.1 pixel for all images.  Finally, data were flux calibrated to
the $V$-band ($\pm 0.1$ mags) using field stars from the UCAC2 catalog
\citep{Zacha2004}.

\begin{figure*}
\centering
\includegraphics[width=17cm]{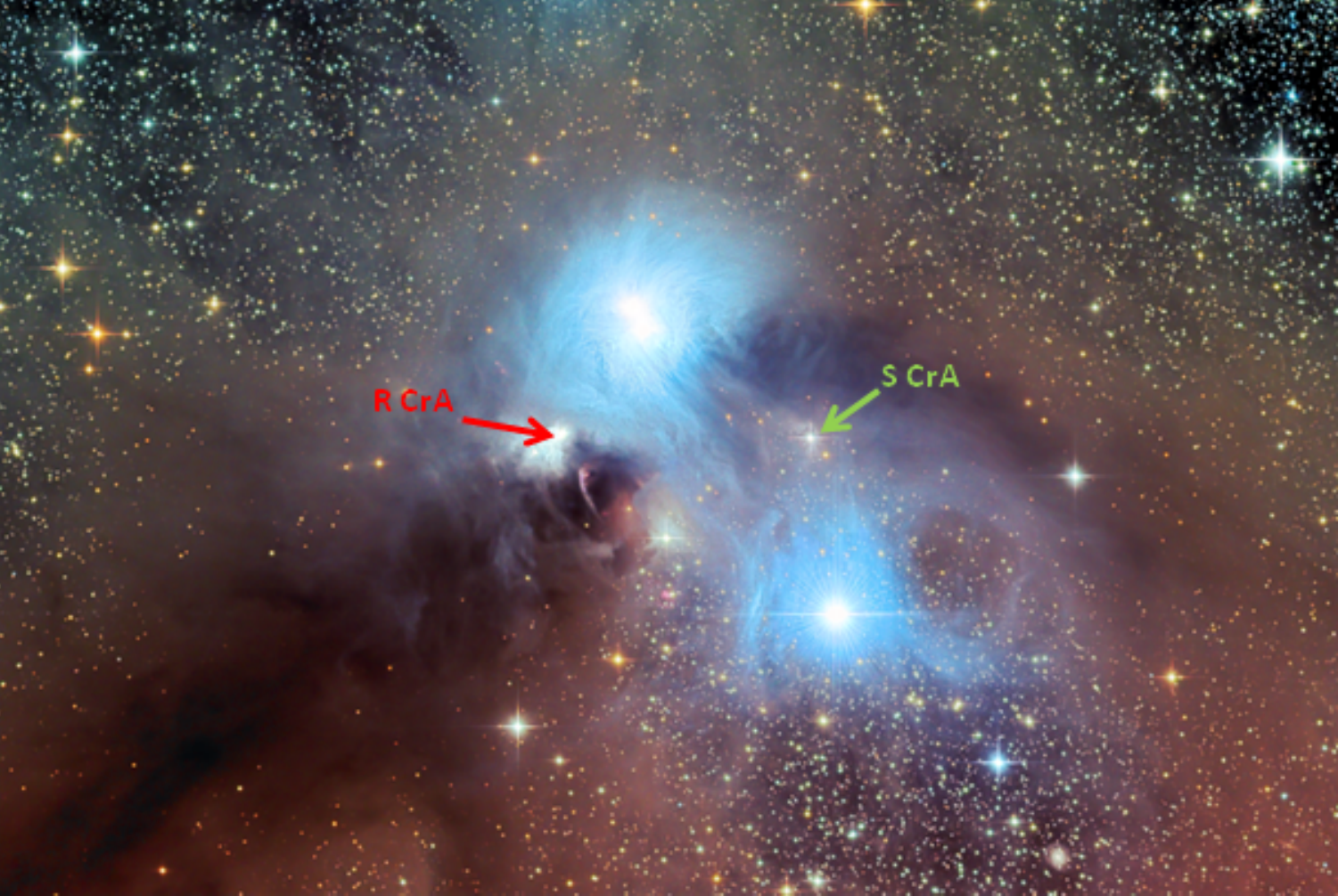}
\caption{Color composite assembled from $B$, $V$, and $R$ images of a
  $47\times 31$ arcmin field of view of the reflection nebula NGC
  6726, with the positions of R CrA and S CrA indicated.  Astronomical
  north is up and east is left.  An animation showing a $30.3\times
  22.1$ arcmin enlargement of the full temporal sequence of
  observations (Table 1) is available online.  Note in the animation
  that time-variable signals are centered on both S CrA and R CrA, and
  that the animation works best when viewed as a continuous loop.}
\label{fig1}
\end{figure*}

\begin{figure*}
\centering
\includegraphics[width=10cm]{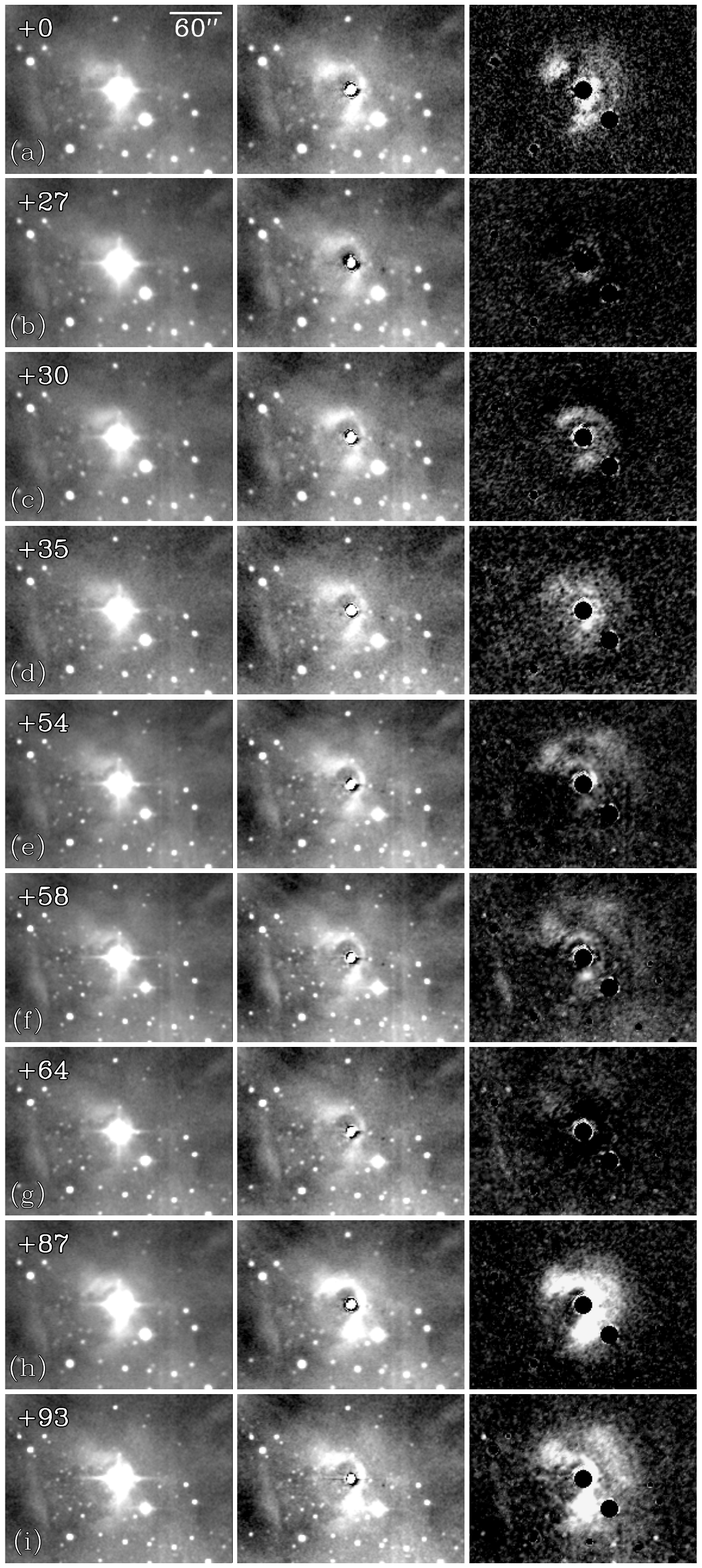}
\caption{Temporal sequences of images of S CrA, showing ({\em left})
  direct images, ({\em middle}) direct images with the flux of S CrA
  removed via PSF subtraction, and ({\em right}) difference-only
  images.  All panels are 270 by 200 arcsec, with north up, and east
  to the left.  The number of days after the first epoch are indicated
  at top-left of each row.}
\label{fig2}
\end{figure*}

Fig.\ref{fig1} shows the reflection nebula NGC 6726 and surrounding field,
with the locations of S and R CrA indicated. A full color animation of that
field is provided as supplementary online material. The time-variable light
echoes are immediately obvious in the animation as they move away from their
sources. A temporal sequence of the field immediately surrounding S CrA is
shown in Fig.\ \ref{fig2}. The echoes can also be faintly seen in the direct
images (left column of Fig.\ \ref{fig2}), and are slightly more apparent
when the point-spread function (PSF) of S CrA has been removed
\citep{Stet1987}, as shown in the middle column.  Since light echoes are
transient phenomena, we employed the PSF-matching and difference-imaging
techniques of \citet{Tomaney1996} and \citet{Suger2005}, combined with the
NN2 algorithm of \cite{Newman2006}, to remove all sources of constant flux
and produce echo-only images for each epoch.  These are shown in the right
column of Fig.\ \ref{fig2}.

An animated close-up view of S CrA subtracted images in which the residuals
from the subtractions have been removed by interpolation is shown as online
material. Expanding concentric rings or ``quasi-rings" are clearly visible
in this animation. As revealed by the animation and in Fig.\ \ref{fig2}, the
echoes have an overall circular morphology but the circles are incomplete;
that is why the term ``quasi-rings" would be more appropriate to describe
them. These expanding echoes are analyzed in the next section.

\section{S CrA Echo Interpretation}

\begin{figure*}
\includegraphics[width=10cm,angle=270]{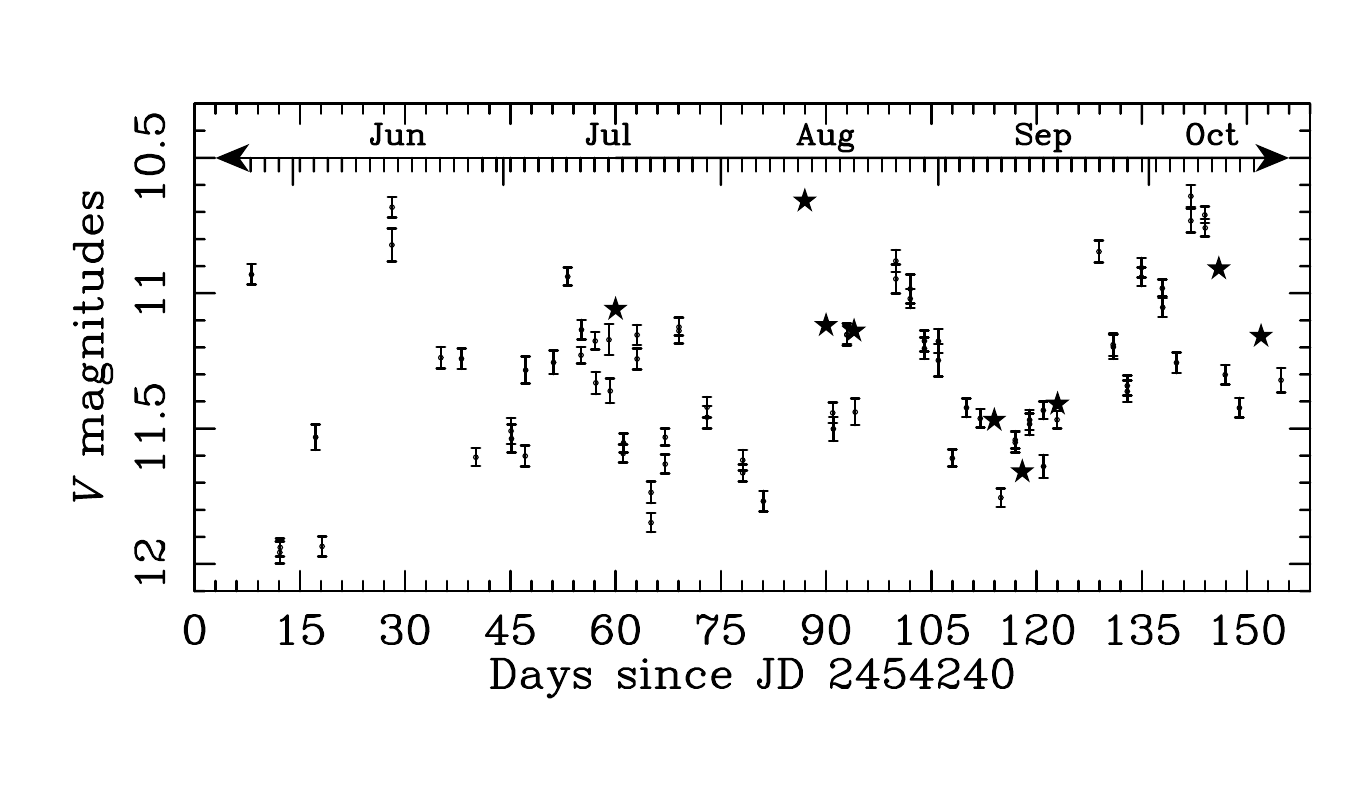} \caption{$V$-band photometry of S
CrA. Points with error bars show data from the All Sky Automated Survey
\citep{Pojman1997} while our data from Table 1 are marked by stars}
\label{fig3}
\end{figure*}

In the typical approach to model nova or supernovae echoes, the light curve
and distance of the source are used to deduce the 3-dimensional structure of
the dust producing the echoes \citep{Suger2003}. Such an approach is not
feasible with our S CrA data. S CrA has been observed as part of the All Sky
Automated Survey \citep{Pojman1997}, however the photometry (Fig.\ref{fig3})
does not have homogenous time sampling, making the epochs of maximum light
and the pulse durations moderately uncertain.  Additionally, there are few
distance measurements to the star, with the best estimates ranging from 120
to 140 pc \citep{Marraco1981,Carmona2007}, as there are no reliable parallax
measurements from Hipparcos \citep[see the
  discussion on this topic in][]{Neu2008}.  In such circumstances, we
instead constrain the above properties by comparing the echoes that
would arise from a test dust geometry to those actually observed. We
initially use guesses for the dates and times of four outbursts and
later on we refine such estimates based on the observed light curve. A
minimum of 4 outbursts are needed for the following reason.  While we
see echoes expanding in the first set of observations, a month later on
we see echoes of smaller angular size, which means that
new echoes have been formed; thus another pulse is needed and so on.

As a first-order model to explain expanding rings, the two simplest dust
geometries are a spherical shell and a planar slab, which are representative
of circumstellar and interstellar media, respectively.  Echoes from a
spherical shell always appear concentric and can only grow as large as the
shell's radius. For a slab, on the other hand, echoes should be observable
to an arbitrary size, and will only appear as concentric, circular rings if
the slab is aligned with the plane of the sky (an unlikely constraint).

Qualitatively, the echoes we observe are concentric at all epochs and end
abruptly around 80 arcsec from S CrA, therefore the spherical shell model is
already favored from these arguments, but for completeness we
explore both the spherical-shell and the planar-slab geometries.

We identified the angular distance of each echo from S CrA (see Table
1), by iteratively fitting Gaussian functions to radial
surface-brightness profiles taken at various position angles
\citep{Suger2005}, avoiding an ``ear''-like filament which
shows brightness variations over time. This ``ear" feature
is stationary and can be explained as an isolated knot with thickness
smaller than $c \Delta t$, where $\Delta t$ is a typical pulse
duration. An isolated knot in the direction perpendicular to the line
of sight would only brighten and fade and no apparent motion would be
seen. Therefore it seems that the ear-like feature is an isolated
overdensity of much smaller size than the dust structure causing the
expanding quasi-rings. Hence we excluded the ear-like signal from the
analysis of the angular size of the expanding rings.

In order to compare the angular distance measurements with theoretical
computations, guesses for the outburst times were inserted into the echo
propagation equations for a spherical cloud model \citep{Crause2005}:

$$\theta =  \sqrt {   2z_0ct - (ct)^2 } /d                 \eqno(1)$$

and a dust slab model \citep{Tylenda2004}:

$$\theta = \sqrt { (1+a)^2(ct)^2 + 2z_0ct } /d         \eqno(2)$$

where $\theta$ is angular radius, $d$ is distance from the observer to the
star, $z_o$ is the distance from the star to the reflecting shell or slab,
$a$ is the tangent of the perpendicular to the dust plane and the line of
sight, $t$ is time from outburst and $c$ is the speed of light.

The outburst times together with $d$ and $z_0$ were iteratively fit by using
a downhill simplex method \citep{Press1992}, in order to minimize the
residuals of the angular radii measurements compared to those resulting from
equations 1 and 2. The results were refined via iteratively-reweighted least
squares \citep{Jeff1987}.

The spherical shell model places the star $121 \pm 29$ pc from earth, with
the shell's average radius $9500 \pm 3800$ AU, and the brightenings occurred
on 9 July, 11 and 25 August, and 06 October 2007 (dates are uncertain by
less than 1 day).  Although the fit to the data is good ($\chi^2 = 0.7$ per
degree of freedom), Fig.\ \ref{fig2} reveals that the echoes observed in
September are very broad, up to 40 arcsec in angular width, requiring very
long pulse durations and/or very thick dust.  Light curves of S CrA from
Fig.\ref{fig3} and the past hundred years \citep{Henden2009} indicate
average brightening last 5--10 days.  Consequently, to generate echoes as
wide as are observed in September, the dust thickness \citep{Suger2003}
would have to be equal to or greater than the dust's distance from the star.
As this is an unlikely constraint, we instead consider whether the echoes
observed in September may in fact be the convolution of light pulses from
early {\em and} late August.  Using the above brightening dates, the echoes
observed in August would be 30 to 40 days old during the September epochs,
placing them at angular radii of 65 to 75 arcsec, exactly where faint echoes
are observed.

When deconvolved into two components, the echoes observed in September have
angular radii of 45 and 63 arcsec (9 September), 52 and 67 arcsec (13
September), and 58 and 70 arcsec (18 September).  With the spherical-shell
model, these revised data yield a distance to S CrA of $138 \pm 16$ pc, a
shell radius of $10000 \pm 1700$ AU, and new brightening dates of 8 July, 10
and 27 August, and 5 October 2007, with a $\chi^2$ per degree of freedom of
0.8.  That our model yields a distance to S CrA consistent with previous
values, and that the epochs of brightening agree with the photometric record
(Fig.\ref{fig3}), confirm that this spherical-shell model is an appropriate
first approximation of the actual dust geometry.

One might argue that echoes coming from a full spherical shell would be
contracting rings after they reach their maximum radii and should have been
seen in our data. However, in order for contracting rings to be detectable,
one would require a considerable amount of backscattering from the dust
particles, which is inconsistent with the properties of the grains that we
have derived (see next section). This is essentially based on the fact that
the surface brightness of the echoes decreases with ring radii, which
implies that the particles are not small and therefore cannot give rise to
considerable backscattering. In conclusion, contracting rings would only be
detectable under very special circumstances that are not met in our case.

Concerning the slab model, it yields two possible solutions.  The first,
with a $\chi^2$ per degree of freedom of 0.7, places S CrA at a distance of
$70 \pm 40$ pc, with the dust inclined by $57\degr\pm 7\degr$ and lying
$6300 \pm 4300$ AU in front of S CrA.  The second model, with a a $\chi^2$
per degree of freedom of 1.0, places S CrA beyond 1 kpc, with the dust
oriented on the plane of the sky and located a few pc in front of the star.
Although these fits are statistically good, the distances to S CrA are
highly discrepant with other estimates.  For the first, or close-slab model,
``L''-type dust (see discussion section for its definition) explains many of
the observed surface brightnesses if the dust density is roughly double that
of the spherical shell model (see discussion section). However, it is
difficult to explain the observed brightness of the largest-radii echoes,
which are at very large scattering angles $(\theta>90\degr)$, at which all
dust types (not just ``L''-type) are inefficient scatterers.  In general,
these echoes are observed to be two or more times brighter than those
expected from dust with large scattering angles.  Besides, the inclination
by $57\degr\pm 7\degr$ would give rise to non concentric rings whose centers
would be displaced several pixels and would therefore be easy to detect, but
this was not the case. In the second, or distant-slab model, one expects
very little change of surface brightness with time, since the echo light is
essentially forward scattering at all times, however this is not observed.
Given the slab models' inconsistencies with distance to the star and dust
scattering properties, we consider that the slab does not appear to be a
good approximation to the 3-dimensional dust structure and we will focus on
the spherical or quasi spherical dust shell.

\begin{figure}
\resizebox{\hsize}{!} {\includegraphics[angle=270]{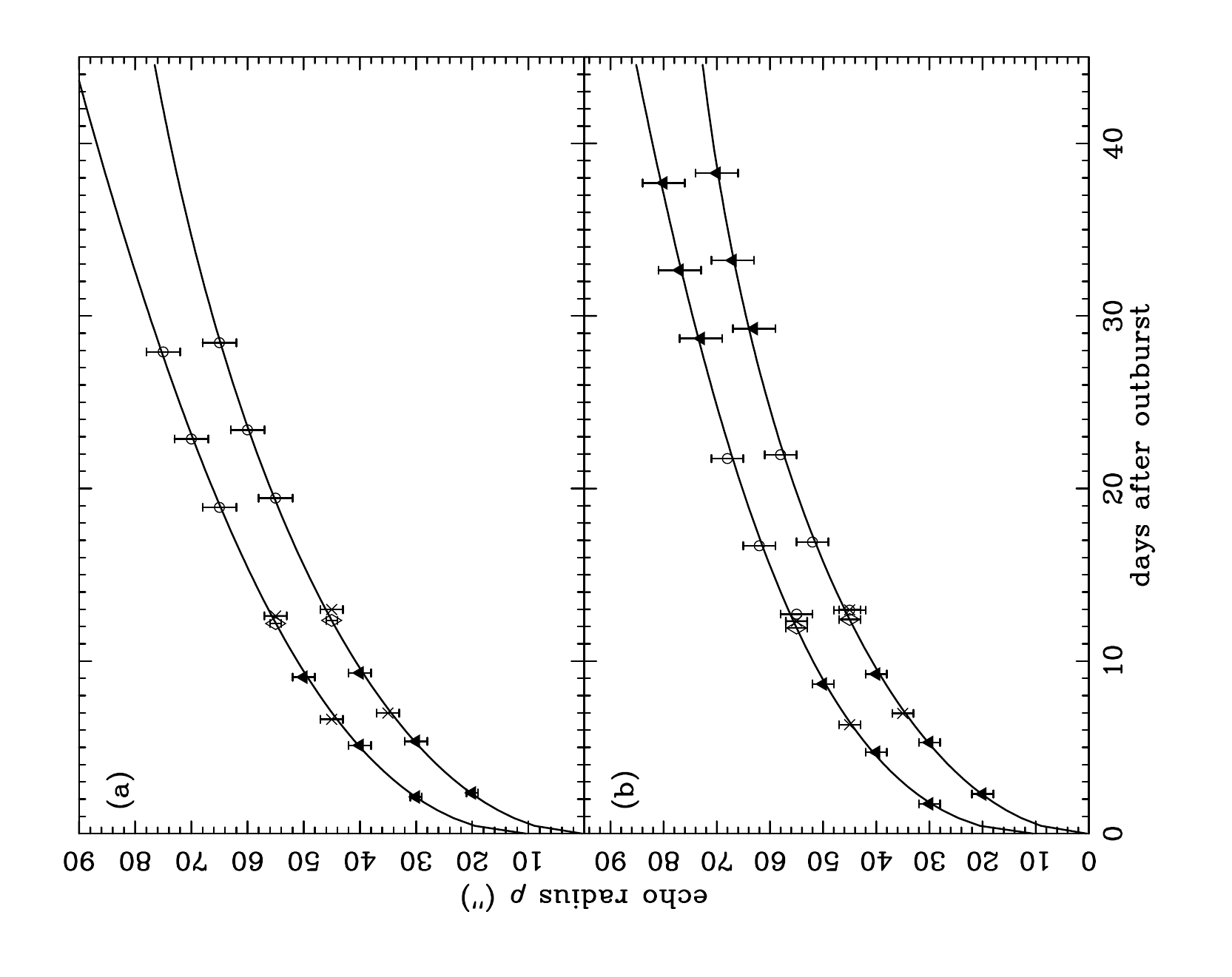}}
\caption{Angular radii of the echoes observed around S CrA versus the time
elapsed since the outbursts that formed them for (a) the data in Table 1,
and (b) revised data assuming two echoes per epoch during September 2007.
The slab model and its associated data are offset 10 arcsec above the
spherical model in both panels. Diamonds denote echoes from July, triangles
from August, circles from September, and crosses from October 2007.}
\label{fig4}
\end{figure}

A plot showing the angular radii measured and the best fits is presented in
Fig. \ref{fig4}

\section{Discussion}

An issue that might require some discussion is the fact that some echoes are
fainter than others of similar age. This can be explained by the different
intensity of the outbursts that give rise to the echoes. In fact, the
farthest (and oldest) echoes from the star were only visible from the
largest outburst (the one that took place around 10th of August 2007).
Echoes as old as 40 days could only be seen from the August 10th outburst
because it was extraordinarily bright (as is demonstrated in the light curve
of \ref{fig3}) compared to the others. Echoes as old as these were not seen in our
data from the other outbursts because a much more sensitive instrument would
have been needed.

Although the echoes appear to have an overall circular or ring aspect, the
flux is generally restricted to the northern half of the image, with a
notable lack of echo flux toward the southeast.  This could imply a lack of
significant circumstellar dust to the south of S CrA, or perhaps
interstellar dust could be selectively blocking the light from the echoes
toward the south east direction. The existence of an outflow in exactly that
direction was shown by \cite{Wang2004} and thus, it is possible that such
outflow is sweeping any dust.

Therefore it must be stressed that the spherical shell that we propose as
the best first order approximation cannot be a complete sphere because we do
not see complete rings but arcs.

What is the origin of the dust?. The most straightforward interpretation is
that such dust is the remains of the stellar envelope, which would still be
dense enough for a star as young as S CrA.  Even though \cite{Carmona2007}
quote an age for S CrA of around 2 Myear and for that age the envelope and
protoplanetary disk should have dissipated already, in their paper
\cite{Carmona2007} implicitly admit that the resolution of their spectra is
too low to detect the photospheric spectrum of this object. High resolution
echelle spectra which show the photospheric spectrum \citep{Appen1986}, give
a lower photospheric temperature, which results in a lower age than
estimated by \cite{Carmona2007}. Most estimates result in $\sim$ 0.5 Myear
or lower. A young age is also indicated by the strong mm and sub-mm dust
emission \citep{Rei1993,Chini2003,Nutter2005,Juvela2009}, the presence of an
HH flow (HH 82) and the high and variable mass accretion rate
\cite[e.g.][]{Walter2005}. Nevertheless, the concentration of dust at $10^4$
AU from the star is somewhat intriguing, and we therefore explore other more
speculative possibilities.

A cloud of dust $10^4$ AU from the star is close to the nominal distance of
the Oort cloud from our Sun \citep{Oort1950}.  Although S CrA is a binary
system, the primary star has a mass similar to the Sun \citep{Carmona2007},
which leads us to speculate that the scattering dust may be an Oort-cloud
analog that is forming, or has formed at least partially around the star.

It is generally believed that the dust disks around young stars disappear
when planets grow from accretion of the disk material. However, it might be
possible that some dust material of the disk is not accreted but dynamically
ejected by the gravitational effect of the forming planets or planetary
embryos, if they are massive enough or if they grow quickly enough. Thus,
planets or planetary embryos have likely formed already from the disk
material of S CrA and therefore it appears possible that at least some dust
from the disk may have been dynamically ejected to large perihelia and with
a range of orbital inclinations.

Giant-planet formation is believed to require1 to 10 Myr
\citep[e.g.][]{Rice2003}. With an estimated age of $\sim$ 0.5 Myr, S CrA may
therefore be actively forming an extra-solar planetary system, but according
to most models, the mass of the planetary embryos would still be far below
the giant planet mass and therefore the proposed dynamical ejection scenario
might not be effective enough.

As noted earlier in the observations and reductions section, the echoes do
not appear as complete rings, which could signal an incomplete or
inhomogeneous shell. This is a point in favor of the Oort-cloud-analog idea.
Since the typical time for complete randomization of the orbital
inclinations of the material ejected from a planetary disk is believed to be
around 500 My \citep{Duncan1987}, one may not expect a fully spherical and
complete Oort cloud from such a young star as S CrA. From Oort-cloud
formation modeling such as the one in \cite{Duncan1987} the expected
geometry of a young forming Oort cloud would be a torus rather than a
sphere. Thus, another possible geometry worth of investigation for the dust
around S CrA is that of a torus. We have briefly tested whether such model
would be consistent with the data, in addition to the slab and spherical
shell already mentioned. Fig.\ref{fig5} shows the regions of space probed by
light echoes; in the first few weeks following a light pulse, echoes from a
spherical shell will appear at all position angles, while those from a torus
will only appear from the northern region.  This leads us to propose that
the echoing structure around S CrA might be compatible with an inclined
torus. Unfortunately, the current dataset is insufficient to adequately
perform detailed 3-dimensional geometric analysis, since the images are not
homogenously sampled in time and the spatial resolution is not high.

\begin{figure*}
\resizebox{\hsize}{!} {\includegraphics[angle=270]{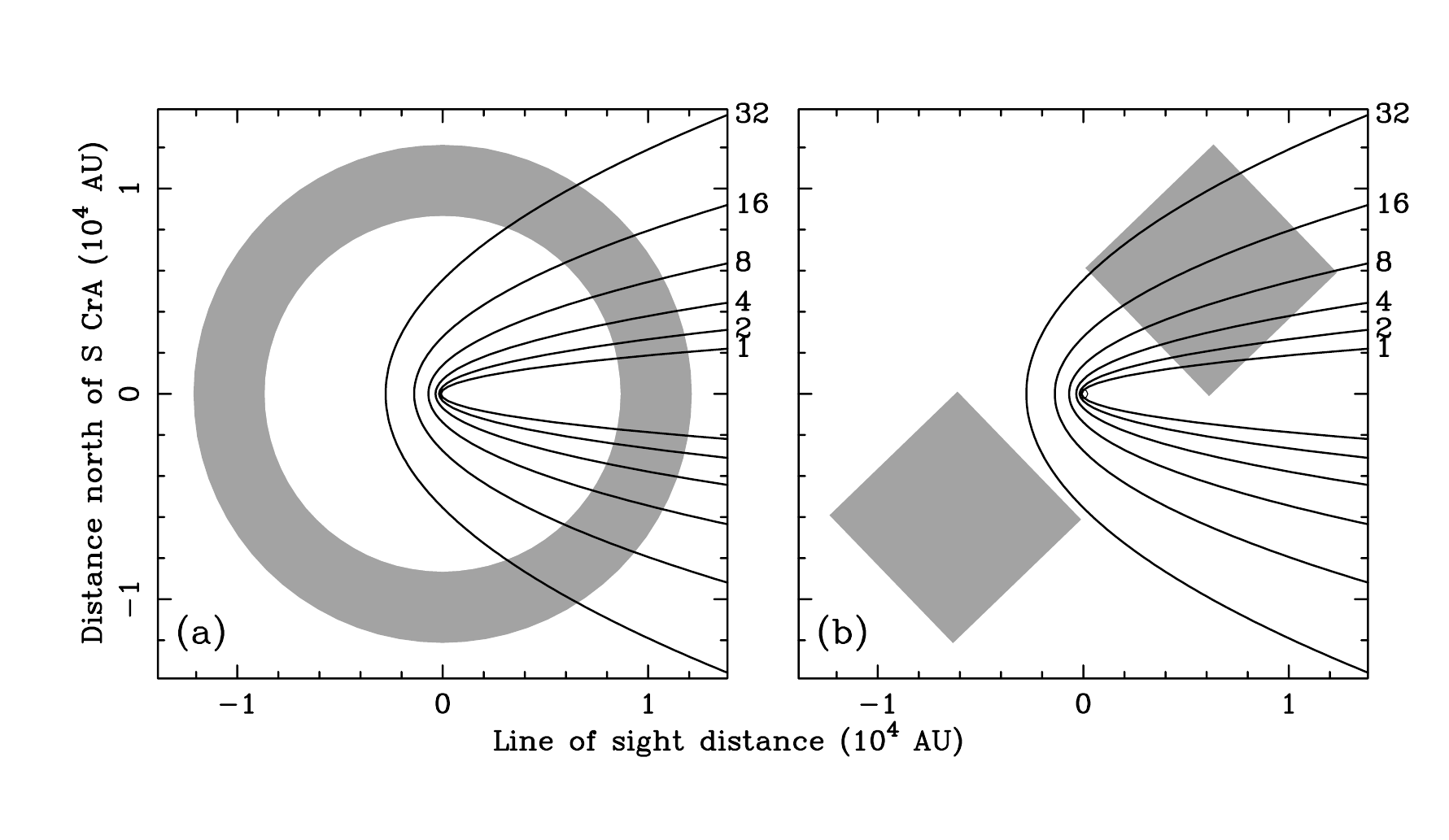}}
\caption{Locations of echoes passing through (a) a spherical shell
  and (b) an inclined torus  of dust.  The number of days after
  maximum light is listed to the right of each parabola.  The abscissa
  gives the distance in front of, or behind, the star, while the
  ordinate shows the distance above (north) or below (south) of the
  star.}
\label{fig5}
\end{figure*}

Even though we are referring to the material that scatters the light pulses
as ``dust", from our observations we cannot determine whether the material
is made of silicates or made of highly volatile condensates (which should be
more properly referred to as ``ice''). Because of the very low temperatures
expected at around 10\,000AU from the star, condensates of highly volatile
gases can exist, not only condensates of H$_2$O. We will continue to refer
to the particles as ``dust", but it must be pointed out that the exact
composition of them cannot be revealed from our observations and therefore
the existence of ice should be taken into account.

Given that the star was only observed in one filter, detail modeling of the
dust size distribution is not feasible, but constraints on the grain sizes
can be derived from the data. If the dust were composed only of small grains
(radius less than 0.01 \micron), then the surface brightness of an echo from
a spherical shell would present small changes with scattering angle. Since
the echoes fade as they move away from S CrA, we conclude that the dust must
include grains with radii $> 0.1$ \micron. Indeed, dust-scattering models
from \citet{Suger2003} suggest that some fraction of the dust must be large,
i.e.\ 1 \micron or larger. Using a large-grain mixture (radii between
0.1-1.0 \micron) of carbonaceous and silicate dust (referred to as ``L"-type
dust), we find the observed surface brightnesses are consistent with a total
dust mass (for a complete spherical shell) of order $2\times10^{-3}$
$M_{\sun}$ and this is likely an upper limit because the shell is not
complete as we already emphasized. A similar mass is found when considering
a cloud of ice particles.  This mass is smaller than that in the
minimum-mass solar nebula \citep{Weiden1977}, suggesting that the observed
amount of dust around S CrA could have originated in its protoplanetary
disk. Estimates of the mass of the current Oort cloud range from several to
100 Earth masses \citep{Weissman1983,Maro1988}, which is considerably lower
than $2\times10^{-3}$ \msun, but one should keep in mind that the Oort cloud
is not made of dust, but of larger bodies, as small grains are not expected
to survive for billions of years due to non gravitational perturbations.

Alternative speculative explanations for the observed dust include mass ejections from
earlier phases of the star's evolution. However, it is well known that
protostellar mass ejections are generally bipolar, and since these cannot
give rise to spheroidal or toroidal geometries, a shell at 10,000 AU would be
a puzzling result from such ejections.

Perhaps a way to test whether these two alternative explanations are correct
would be to measure the amount of gas that is present at 10,000 AU. In the
Oort cloud analog scenario one would expect no gas at all, whereas in the
other two scenarios one would expect gas to dust ratios typical of the
interstellar medium. Even though there is no concluding evidence in the
literature, radio maps that trace CO gas show little or no gas around S CrA
\citep{Harju1993} whereas there is emission in the submillimetric and
millimetric ranges coming from dust as already mentioned.

We therefore speculate that the dust producing the observed echoes is
perhaps a young Oort-cloud analog, even though the most straightforward
interpretation for the dust shell is the remains of S CrA's dust envelope,
given S CrA's young age. As noted earlier, in order for a young Oort-cloud
analog to form, the star system would require at least one massive planet
forming inside the protoplanetary disk, to actively eject planetesimals and
debris to larger orbital radii. In that case, an intriguing implication is
that light echoes around young stellar systems could be used to indirectly
detect the presence of jovian or massive planets.

\section{Conclusions}

We have detected not just one, but multiple echoes from light pulses of very
young stars. Such echoes are an impressive document and indicate that
similar echoes might be detectable around other young nearby stars or in star
forming regions. This has good potential as a tool to study the environment
around young stars and the process of planet formation in those systems.

The echoes surrounding S CrA are more confined and are much easier to
interpret than those coming from R CrA although the latter are more intense.
We have presented a preliminary analysis of the echoes surrounding S CrA.
From the information contained in the angular distance of the echoes versus
time from outbursts we have been able to derive basic properties of the dust
that is scattering the light-pulses. The dust must be concentrated in a
region around 10\,000 AU from the star and the best geometry of the dust in
terms of reproducing the observations is an incomplete spherical shell or an
inclined torus; a planar slab geometry appears to be ruled out by the
observations, but it would require dust at a similar distance in order of
magnitude. The best fit to the echo radii gives a distance from Earth to S
CrA of 128 $\pm$ 16 pc in the case of the spherical shell distribution,
which is consistent with previous rough estimations of the distance to S CrA
\citep[see the distance discussion in][]{Neu2008} whereas the dust slab
geometry gives a distance to S CrA that cannot be reconciled with other
observations in the literature. The derived distance implies that these are
the closest echoes to the Earth that have ever been recorded.

Concerning the information extracted from the surface brightness of the
echoes, the mass of the dust shell required in order to reproduce the
observed echoes surface brightness is in the order of ${ 2\times10^{-3}}$
\msun and must be in the form of particles larger than 0.1$\mu m$ to explain
the fading of the echoes as they grow in angular size. Such dust structure
is probably the remnant of the envelope typical of very young stars. But
because of the concentration near 10000 AU, we speculate that the dust
particles might have been dynamically ejected from the star's disk through
the action of forming planets and ended up where they are now in a similar
way as planetesimals have been ejected from the solar system to form the
Oort cloud.

We must caution the reader that we have presented just a first order model
to explain the echoes around S CrA. A well sampled lightcurve of SCrA and
high spatial resolution images would be needed in order to derive the
detailed 3-dimensional structure of the dust surrounding S CrA.

%%%%%%%%%%%%%%%%%%%%%%%%%%%%%%%%%%%%%%%%%%%%%%%%%%%%%%%%%%%%%%%%%%

\begin{acknowledgements}

We are grateful to M. Fern\'{a}ndez, M. Osorio and G. Anglada for very
helpful discussions. This research was partially supported by Spanish grants
AYA2005-07808-C03-01, PCI2005-A7-0180, AYA2008-06202-C03-01, P07-FQM-02998
and European FEDER funds. B.E.K.S. was supported, in part, by HST grant
GO-10607. We acknowledge with thanks the variable star observations from the
AAVSO International Database contributed by observers worldwide and used in
this research. We also acknowledge the use of ASAS database. RD acknowledges
financial support from the MICINN (contract Ram\'on y Cajal). We also
acknowledge the fruitful comments from the anonymous referee.

\end{acknowledgements}

\newpage

\begin{table*}
\caption{Observing Log }
\label{tbl1}
\begin{tabular}{lccccc}
\hline \hline \vspace{0.1cm}
 Date & UT & Exp time (s) & $V$ Mag$^{\mathrm{a}}$ & Angular Distance$^{\mathrm{b}}$  & Surface brightness$^{\mathrm{c}}$  \\
 \hline

2454298.62 & 2007 07 17  02:49 & 1800 & 11.1 & 45 & 2.0\\
2454325.59 & 2007 08 13  02:12 & 1800 & 10.7 & 20 & 1.2\\
2454328.57 & 2007 08 16  01:46 & 1800 & 11.1 & 30 & 2.7\\
2454332.54 & 2007 08 20  00:57 & 1800 & 11.1 & 40 & 1.6\\
2454352.56 & 2007 09 09  01:33 & 2400 & 11.5 & 55 & 2.0\\
2454356.52 & 2007 09 13  00:34 & 2400 & 11.7 & 60 & 1.5\\
2454361.58 & 2007 09 18  01:52 & 3000 & 11.4 & 65 & 0.7\\
2454384.52 & 2007 10 11  00:32 & 2400 & 10.9 & 35 & 5.0\\
2454390.52 & 2007 10 17  00:38 & 3000 & 11.2 & 45 & 3.4\\
\hline
\end{tabular}
\begin{list}{}{}
\item[$^{\mathrm{a}}$] Typical uncertainties are 0.2 mags.
\item[$^{\mathrm{b}}$] Given in arcsec; typical uncertainties are 2\arcsec.
\item[$^{\mathrm{c}}$] Given in units of $10^{-19}$ \ecsaa; typical
uncertainties are $10^{-20}$ \ecsaa.
\end{list}
\end{table*}

\section{On-line Material}
\begin{figure}[h!]
\resizebox{\hsize}{!} {\includegraphics[width=8cm]{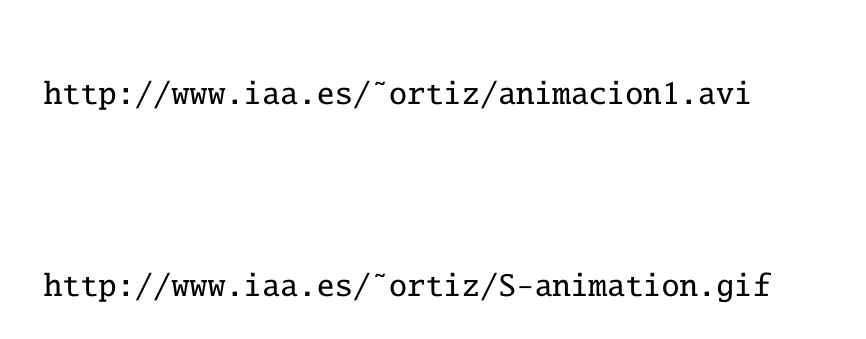}}
\label{fig5}
\end{figure}

\end{document}